\documentstyle[aps,prl,graphicx]{revtex}

\begin{document}
\title{Further evidence for narrow exotic low mass baryons.}

\author{
B. Tatischeff\thanks{E-mail : tati@ipno.in2p3.fr}}


\address{
Institut de Physique Nucl\'eaire,CNRS/IN2P3, F--91406 Orsay Cedex,
France}

\maketitle
\vspace*{1cm}
\begin{abstract}
Although narrow low mass baryonic structures, observed mainly in SPES3 (Saturne)
data, were not confirmed in recent experiments using lepton probes, their
existence is confirmed in other data, where hadronic probes were used.
\end{abstract}

\vspace*{4.mm}
\hspace*{2.cm}PACS numbers: 13.75.Cs, 14.20.Gk  

\vspace*{5.mm}
\hspace*{4.mm}
Low mass narrow baryonic structures have been observed in different
experiments over many years. In spite of the fact that the corresponding peaks
have been extracted from some data, and with reasonable statistics, they
were often considered with scepticism. Such scepticism may be justified by the
absence of confirmation in spite of different searches \cite{zol} \cite{kol},
and also by the importance of these observations, especially since the lowest
masses of these narrow exotic baryons are below the pion production mass.
These results therefore need to be confirmed. Many reactions can be used,
provided the statistics are large enough and the resolution
good ($\sigma\approx$~5~MeV). An incident beam energy of a few GeV
allows to excite these structures over a large range
1$\le$M$\le$2~GeV while keeping the resolution good enough.
It seems also that these observations are more readily observed with
hadronic, as opposed to leptonic probes \cite{kol}.\\
\hspace*{4.mm}
Although new dedicated experiments are needed, it is possible to look at these
states in already published data.
More often, if not always, the authors have ignored the idea of associating
narrow discontinuities
of their spectra with such physics. In some cases the real physical effect
could not be ascertained, generally due to the insufficient knowledge of the
acceptance of the detection device. Since the measurements were not performed with
the aim of looking for small peaks or shoulders, the statistics were generally
too small for the purposes discussed here.\\
\hspace*{4.mm}
It is the aim of this paper to take a closer look at some previously published
results and to conclude on the existence of narrow exotic low mass baryons.\\ 
\hspace*{4.mm}
The first published paper, advocating the existence of exotic narrow low
mass baryons, analyzed the missing mass data from pp$\to$p$\pi^{+}$X
reaction \cite{bor1}. The experiments were performed using the SPES3 facility
at Saturne. Due to the good resolution and high statistics,
three structures at M$_{X}$=1004, 1044, and 1094~MeV were extracted. The
experimental widths ($\sigma$) of these structures vary from 4 up to 15
MeV. The physical widths are low; indeed the two first structures can only
de-excite
through electromagnetic interactions. These baryons were tentatively
associated with two colored quark clusters: either $q-q^{2}$ or
$q^{3}-q\bar{q}$. While all three previously quoted structures, are
observed at most angles and energies, narrow structures are also observed at
higher masses, but are lightly excited and therefore more rare. Using data from
pp$\to$p$\pi^{+}$X and pp$\to$ppX reactions studied at SPES3 \cite{bor2},
and using the spectra of $\gamma$n$\to$~p$\pi^{-}\pi^{0}$ from MAMI \cite{zab}
and the spectra of $\gamma$p$\to\pi^{+}$n from Bonn \cite{dann}, structures
were extracted at the following masses: M=1136, 1173, 1249, 1277, and 1384~MeV
\cite{bor2}. Another structure at M=1339~MeV was also extracted, but with a lower
confidence level, see fig.46 of \cite{bor2}.\\
\hspace*{4.mm}
The study of the mass region 1470$\le$M$\le$1695~MeV is underway and the
results will be published later \cite{bort}. The situation in this mass range
is complex since many narrow structures are observed in the total
baryonic spectra and in spectra associated with particular de-excitation
modes: p$\pi^{0}$ and p$\eta$. In the M$_{pX}$ spectra, narrow structures
are extracted at the following mean mass values: M=1479, 1505, 1533, 1577,
1639, 1659, and 1669~MeV. In the n$\pi^{+}$ de-excitation mode, structures
are extracted at the following mean masses: 1505 and 1542~MeV.
In the p$\pi^{0}$ de-excitation mode, structures are extracted at the
following mean masses: M=1517, 1533, 1564, 1601, and 1622~MeV. In the p$\eta$
de-excitation mode, structures are extracted at the following mean masses:
M=1505, 1517, 1564, 1577, 1639, and 1659~MeV. A peak at a mean mass of
1554~MeV, is less well defined, but extracted at close mass in M$_{n\pi^{+}}$,
M$_{ps\pi^{0}}$, and M$_{pf\eta}$.\\
\hspace*{4.mm}
The mass range 1720$\le$M$\le$1790~MeV was
also studied at SPES3, using the same reactions as those given before, but
at higher incident energy. Two structures at M=1747 and 1772~MeV were extracted
\cite{bor3}.\\
\hspace*{4.mm}Structures were also observed at masses lower than 1~GeV.
The pd$\to$ppX
reaction was studied at the Linear Accelerator of INR (Moscow) and peaks
were extracted in the missing mass M$_{X}$ at 966, 986, and 1003~MeV
\cite{fil1}.
Several narrow structures were also extracted in this mass region, from
SPES3 spectra. Preliminary results were presented at the Baldin's Conference
held at Dubna (2002) \cite{bal2}.\\  
\hspace*{4.mm}Narrow structures can also be extracted from data previously
studied for different motivations. A few spectra, studying the p(d,d')X
reaction, were measured at SPES4 (Saturne)
\cite{ban1} thirty years ago.
At that time, the SPES4 detection system was composed of several pairs of
scintillation counters, without drift chambers. The resolution was therefore
poor and consequently the binning was also poor. Among five spectra at
different incident energies
and angles, in only one spectrum corresponding to p$_{inc.}$=2.95~GeV/c and
$\theta_{lab}$=4.6$^{0}$, a small, but statistically well defined shoulder,
is observed at M$_{X}\approx$1350~MeV. This mass is to be compared to the mass
M=1339~MeV extracted from the SPES3 data.\\
\hspace*{4.mm}More recently, the analyzing powers of inelastic {\it dp}
scattering were
measured using the SPES4-$\pi$ setup, in the energy of the $\Delta$ and Roper
resonance's excitation \cite{mali}. Here the statistics are low.
Two spectra show the number of events versus the missing mass. In the first
one, corresponding to the p(d,d')n$\pi^{+}$ reaction, no structure can be
extracted. In the second one, see Fig.~8(b) of \cite{mali}, although the
statistics are poor, a peak can be
extracted at M$\approx$1140~MeV, a mass very close to M=1136~MeV, where a
narrow peak was extracted from the SPES3 data.\\
\hspace*{4.mm}
Precise spectra of the  p($\alpha$,$\alpha$')X reaction were obtained at SPES4
(Saturne) in order to study the radial excitation of the nucleon in the
P$_{11}$(1440~MeV) Roper resonance. The measurements were done using a
T$_{\alpha}$=4.2~GeV incident beam. The spectrum at
$\theta_{\alpha'}$=0.8$^{0}$
was published in \cite{mor1} and the spectrum at $\theta_{\alpha'}$=2$^{0}$
was published
in \cite{mor2}. A first large peak around M$_{X}\approx$1130~MeV
($\omega\approx$240~MeV), was associated
with the projectile excitation, and a second large peak around
M$_{X}\approx$1345~MeV ($\omega\approx$510~MeV), was associated
with the target excitation. Above them lie narrow peaks, defined
by a large number of standard deviations, since the highest channel at
$\theta$=0.8$^{0}$ contains approximately 2.5$\times$10$^{4}$ events
 (see Fig.~1). These structures were not discussed by the authors.\\
\hspace*{4.mm}
A short discussion is useful, to describe the detection system in order to
ascertain the physical reality of these structures. It was previously
described in \cite{bedj}. The six planes of the two drift chambers, used
to analyze the detected particle momenta, could not be at the origin of
significant particle losses. Nor could the trigger scintillator planes at the
focal plane, since each trajectory was covered by two counters with partial
overlap. There was one trigger plane at the intermediate focus, consisting of
an hodoscope of 12 contiguous scintillator
counters of 2.16~cm width. The spectrometer total momentum acceptance was
6$\%$. Would there have been `holes' in the data due to particles passing
between two scintillators of the hodoscope located in the intermediate focus,
these losses would
be narrow. Indeed the ratio of `hole' over full counting will be $\approx$~4$\%$.
The `holes' will be separated by $\Delta$p/p=0.5$\%$
($\Delta$M$_{X}\approx$~28~MeV)
around M$_{X}$=1227~MeV at $\theta$=2$^{0}$. Such a behaviour is not observed
in the data. Calibration spectra, obtained with unfocalized beam, were
totally flat \cite{mor3}. Moreover, spectra from different reactions, which have
been studied
with the same detection system, displayed continuous behaviours. For example
the spectra of p($^{3}He,t)\Delta$ reaction are smooth, but with poorer
statistics \cite{con}.
 The resolution of the SPES4 detection system is
 $\Delta$p/p=0.22$\times$10$^{-4}$,
allowing fully to exploit the SPES4 performances
$\Delta$p/p=2.2$\times$10$^{-4}$
\cite{bedj}. The corresponding resolution for M$_{X}$=1227~MeV, in the data
obtained at $\theta$=2$^{0}$, T$_{\alpha}$=1420~MeV, is
$\Delta$M$_{X}$=8~MeV. The data are binned into $\approx$5~MeV/channel in
M$_{X}$, and almost all peaks are defined by a few channels. The spectra are
composed of five spectra measured with different spectrometer fields. Each
spectra was slightly renormalized by a factor less than 2, to correct for
different monitor counts. The statistical errors could not be larger than a
factor of 2, from the error extracted using the given counts \cite{mor3}.
The error bars are therefore multiplied by this factor 2.\\
\hspace*{4.mm} The data from the p($\alpha,\alpha$')X reaction 
\cite{mor1}, \cite{mor2} were used, and the masses of the structures extracted.
Fig.~1
shows the yield at $\theta$=0.8$^{0}$, their error bars and the extracted
structures. The empty circles are the published number of events versus the
energy loss. They correspond to the scale. The full squares
show the data in a scale enhanced by a factor 7, and the full circles show
the data enhanced by a factor 15. The masses corresponding to the extracted
narrow
structures are shown in table~1. All peaks seen in SPES3 data, and also in 
SPES4 data, have been introduced in the discussion. At this very forward angle,
the incident beam enters the SPES4 spectrometer, preventing measurements at
momenta close to the elastic scattering limit (low $\omega$), and therefore
preventing a possible confirmation of the lower mass structure M=1004~MeV,
seen in SPES3 data. Two peaks at M=1394~MeV and M=1428~MeV, extracted from
SPES4 data, were not seen in SPES3 data, since such mass range was not
studied in the SPES3 experiments.\\
\hspace*{4.mm}Fig.~2 shows the corresponding yield at $\theta$=2$^{0}$. Here
again the empty circles show the published number of events and
the full circles show the data enhanced by a factor 10, while the full
squares show the data enhanced by a factor 2.4. At this angle, nearly all
low mass
structures, observed previously at SPES3, are present. One peak at M=1277,
extracted from SPES3 data, is not observed in $\theta$=2$^{0}$ SPES4 data,
but is observed at M=1268~MeV in the $\theta$=0.8$^{0}$ SPES4 data.
The peaks at M=1394~MeV and M=1428~MeV, not observed in the SPES3 data, are
also not seen in these SPES4 $\theta$=2$^{0}$ data.
Above M=1470~MeV, the same
complexity is observed in the p($\alpha,\alpha$')X data, see Fig.~2, as 
was observed in the SPES3 data. Indeed, the energy loss range
830$\le\omega\le$1150~MeV, corresponds at $\theta$=2$^{0}$ to the mass range
1470$\le$M$\le$1590~MeV.\\
\hspace*{4.mm}The widths ($\sigma$) of these peaks observed in the SPES4
data vary from 10 to 20 MeV. Fig~3 shows for both angles, the comparison
between narrow peak masses, extracted from SPES3 and SPES4 data.
The straight line corresponds to the same masses, and almost all points are
located along this line. Otherwise, they are located near the line.
 The data points are empty for less
well defined peaks, and mainly concern unpublished SPES3 data
in the 1480$\le$M$\le$1600~MeV mass range.\\
\hspace*{4.mm}
The positions of the structures extracted from p(d,d')X spectra, and especially
 from p($\alpha,\alpha$')X spectra,
correspond, with high accuracy, to the positions observed from SPES3 data in
various reactions, mainly pp$\to$p$\pi^{+}$X and pp$\to$ppX. Such a concordance
is also observed for the lowest masses located below the pion production
threshold. This agreement is obtained using data, studied for different
motivations and previously published, obtained by different physicists, using
different experimental set-ups, and different incident beams and reactions.
This result contributes strongly to the question of the real existence
of these exotic baryons.\\
\hspace*{4.mm}These baryons were tentatively associated with $q^{2}-q$ and
$q^{3}-q{\bar q}$ quark clusters \cite{bor2}. Since they are observed in
p($\alpha,\alpha'$)X reactions, they must all have isospin 1/2. Fig.~46 of 
\cite{bor2} shows that such an isospin attribution is compatible with the mass
formula and parameters used in \cite{bor2}.\\
\hspace*{4.mm}
I would like to acknowledge J.L. Boyard for the discussions concerning the
properties of the SPES4 detection system, M. MacCormick for help in
writing the paper in English, and H.P. Morsch for informations concerning the
SPES4 experiment and for providing the numerical values of the
p($\alpha,\alpha$')X spectra at $\theta$=0.8$^{0}$.

\begin{table}
\caption{Masses (in MeV) of narrow exotic baryons, observed previously in SPES3 data
and extracted from previous p($\alpha,\alpha$')X spectra measured at SPES4
[13] [14].}
\label{Table 1}
\end{table}
\begin{table}
\begin{tabular} [h]{c c c c c c c c c c c c c} 
\hline
SPES3 mass&1004&1044&1094&1136&1173&1249&1277&1339&1384&&&1479\\
pic marker&(a)&(b)&(c)&(d)&(e)&(f)&(g)&(h)&(i)&(j)&(k)&(l)\\
SPES4 mass 0.8$^{0}$&&1052&1113&1142&1202&1235&1262&1322&1370&1394&1428&1478\\
SPES4 mass 2$^{0}$&996&1036&1104&1144&1198&1234&&1313&1370&&&1477\\
\hline
\hline
SPES3 mass&1505&1517&1533&1542&(1554)&1564&1577&&&&\\
pic marker&(m)&(n)&(o)&(p)&(q)&(r)&(s)&&&\\
SPES4 mass 2$^{0}$&1507&1517&1528&1544&1557&1569&1580&&&&\\
\hline
\end{tabular}
\end{table}

\begin{figure}[!h]
\begin{center}
\scalebox{1}[1]{
\includegraphics[bb=1 1 530 530,clip=]{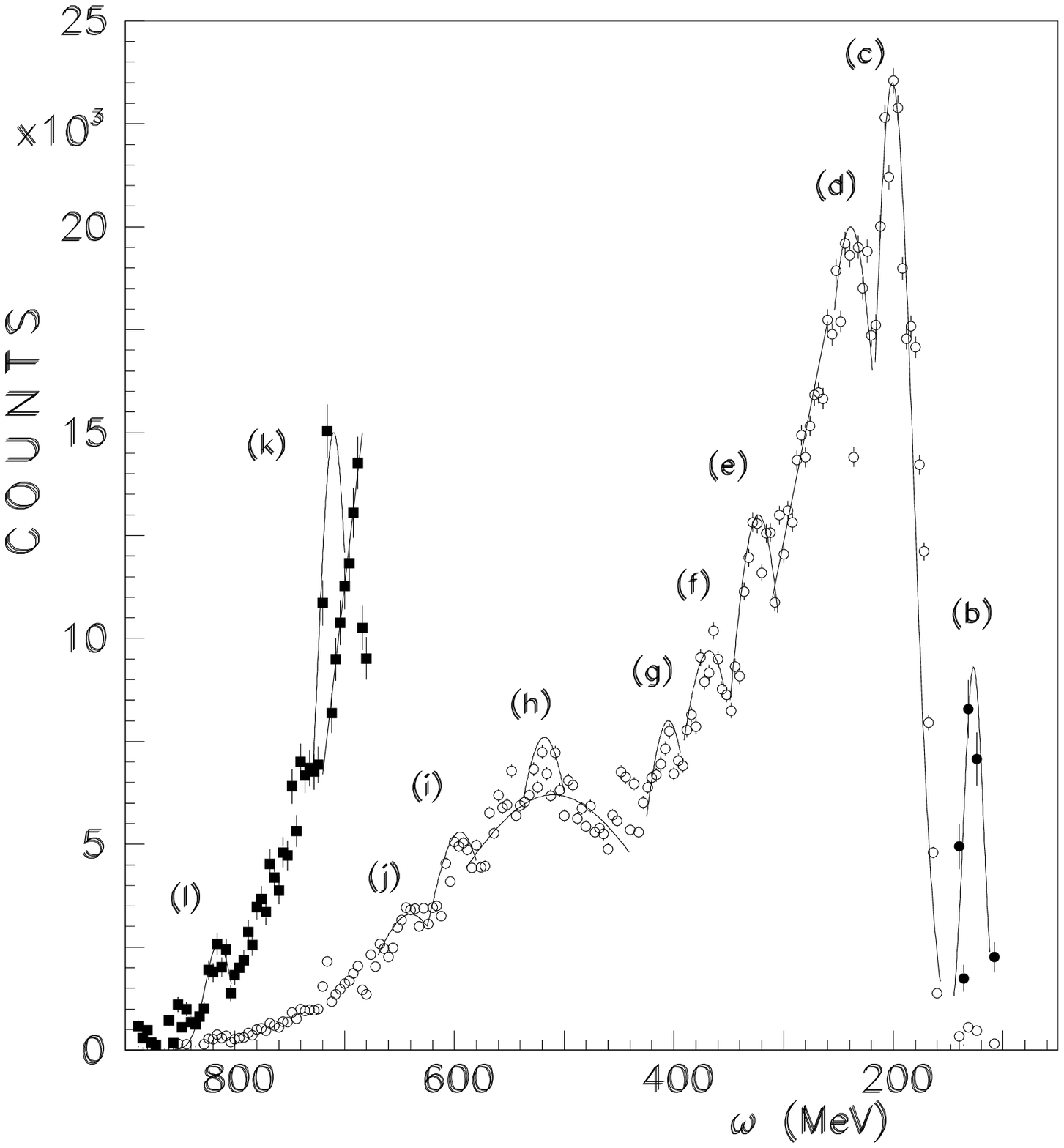}}
\caption{Spectra of the p($\alpha,\alpha'$)X reaction studied at SPES4
(Saturne) with T$_{\alpha}$=4.2~GeV and $\theta$=0.8$^{0}$ [13].}
\label{fig1}
\end{center}
\end{figure}

\begin{figure}[!h]
\begin{center}
\scalebox{1}[1]{
\includegraphics[bb=1 1 530 530,clip=]{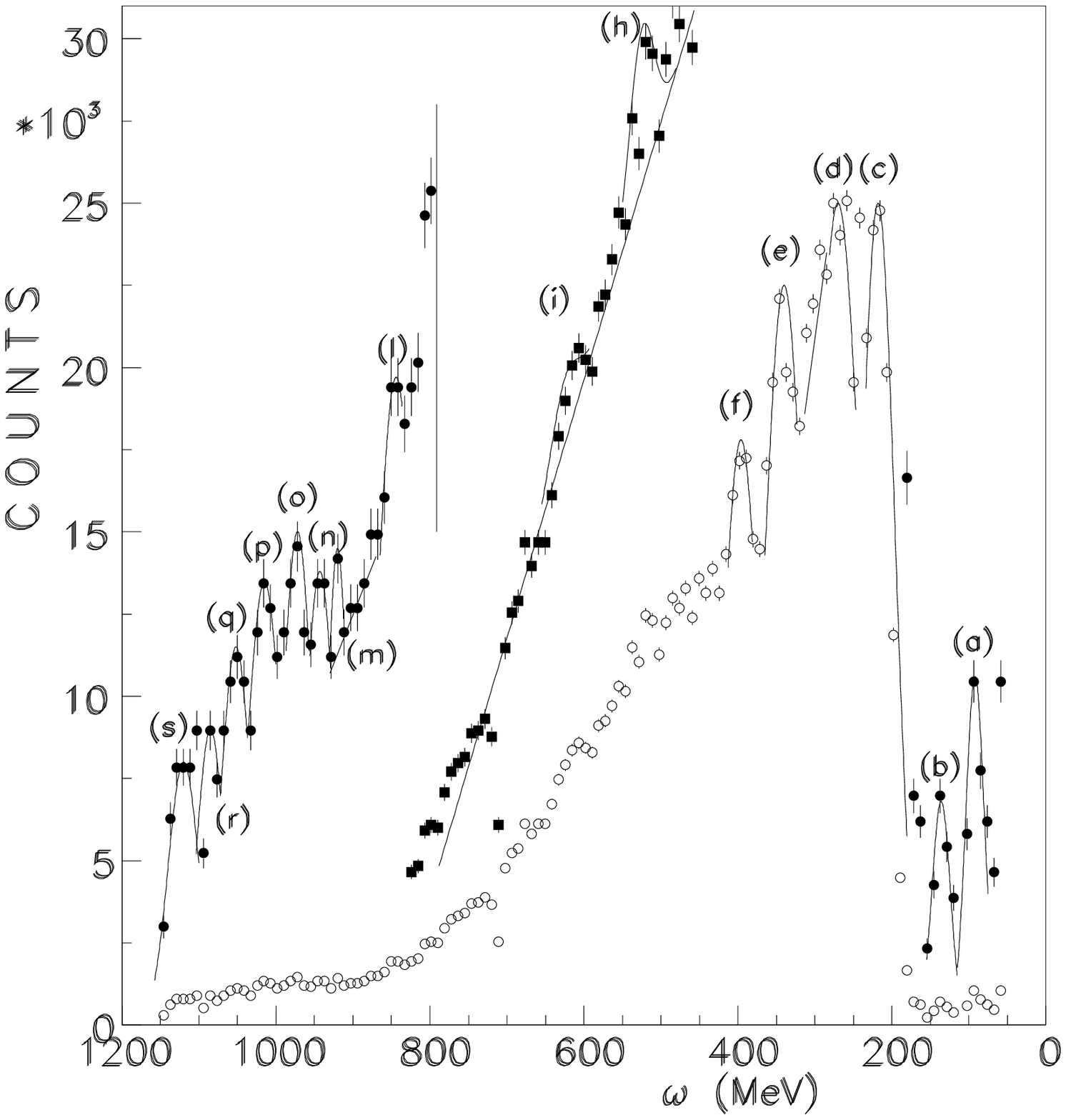}}
\caption{Same caption as Fig.~1, but obtained at $\theta$=2$^{0}$ [14].}
\label{fig2}
\end{center}
\end{figure}

\begin{figure}[!h]
\begin{center}
\scalebox{1}[1]{
\includegraphics[bb=1 1 530 530,clip=]{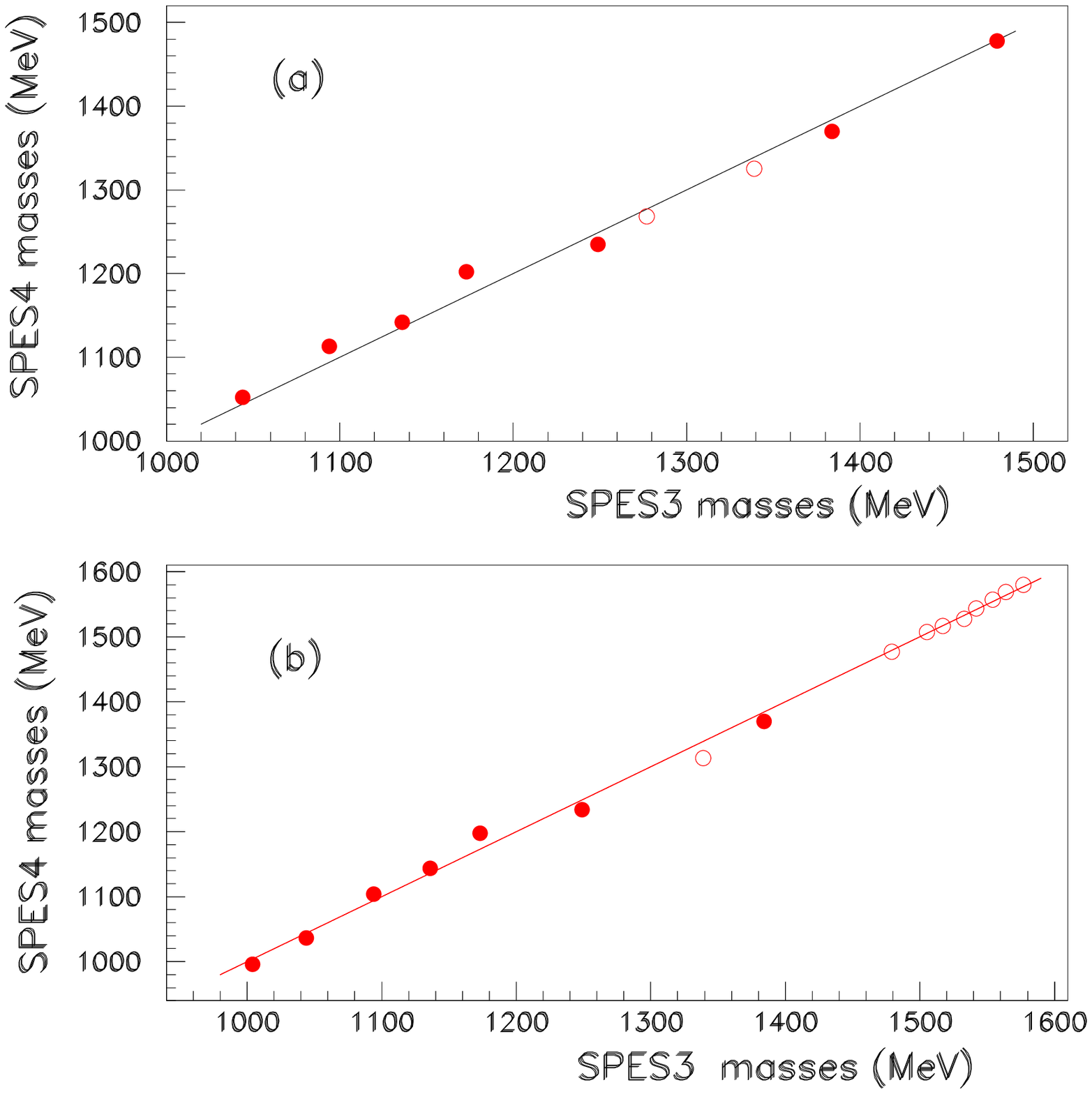}}
\caption{Comparison between masses of narrow baryons extracted from
SPES3 and SPES4 data. Inserts (a) and (b) correspond respectively to
$\theta$=0.8$^{0}$ and $\theta$=2$^{0}$.}
\label{fig3}
\end{center}
\end{figure}

\end{document}